# Disease-level drug approval forecasting from trial-agnostic public data

Samir Townsley[1,2], Pagkratios Tagkopoulos[1,2], Ilias Tagkopoulos[1,2*]

[1] Department of Computer Science and Genome Center, University of California, Davis, USA

[2] USDA AI Institute for Food Systems (AIFS), Davis, CA, USA

*Corresponding author:

Email: itagkopoulos@ucdavis.edu

## Abstract

Drug development is an expensive and lengthy process requiring developers to select drugs for human trials and diseases to target. Computational tools that can estimate before testing which drug-disease candidates are most likely to reach FDA approval could reduce development time and costs through early program advancement and disease prioritization. Yet existing approaches predict outcomes at a single level or stage or do not distinguish between disease outcomes for the same drug. We ask (1) whether FDA approval can be forecast for a drug-disease pair using only pre-clinical trial information, and (2) whether these forecasts can help prioritize diseases for the same drug. We built a cohort of 14,134 drug-disease pairs by reconstructing programs from ClinicalTrials.gov, extracting disease-level approvals from FDA label documentation, and enriching each pair with molecular, disease, ADMET, target, and pathway features. We divided the cohort into pre- and post-2019 programs and trained a classifier on pre-2019 data to predict FDA approval. On the post-2019 cohort (2,637 programs), the model would have prioritized a program that ultimately gained approval over one that did not approximately three out of four times. For drugs tested across multiple diseases, selecting the model's highest-ranked disease would have identified an ultimately approved disease 61.3% of the time, compared with 40.7% under random selection. These results show that public pre-trial information contains signal for forecasting FDA approval at the drug-disease level and can help prioritize diseases for the same drug. We provide a public web tool for SMILES- and ICD-10-based drug-disease approval forecasting.

## Introduction

Bringing a drug to market is slow and expensive. Drug developers and investors commit billions of dollars to human trials to determine whether a candidate will ultimately succeed. Including the cost of failed programs, one estimate places capitalized R&D spending at roughly US$2.6 billion per approved drug.[1] While drug discovery often emphasizes efficacy, successful drug development depends on multiple dimensions, including safety, toxicology, pharmacokinetics/bioavailability, and commercial viability.[2] FDA approval specifically requires clinical evidence of safety, efficacy, and a favorable benefit-risk balance for the intended use.[3] Historical analyses of drug success rates show that, despite the required investment of time and capital, only an estimated 10-14% of drugs that enter clinical testing ultimately satisfy all of the criteria to reach FDA approval.[4,5] Tools that can reliably forecast approvals have the potential to eliminate low probability candidates earlier in the drug development process, leading to improved success rates and substantial cost and time reduction. While historical trends show Phase 2 trials to be the biggest bottleneck in clinical development, substantial attrition occurs across all clinical trial phases, making approval forecasting an end-to-end challenge.[5]

We examine whether eventual FDA approval can be forecast before the start of clinical trials for a specific drug-disease program. Such a forecast can support two related decisions: whether a particular drug-disease program should be considered for human testing and, when multiple diseases are under consideration for the same drug, which should be prioritized. Note that these are distinct from several frequently explored questions in literature including (1) Should a particular drug enter clinical trials for any disease? (2) Will a given clinical trial meet its primary endpoints? (3) Will an ongoing drug development program lead to an FDA approval? The relevant unit of prediction in our question is the drug-disease pair: a molecule may succeed for one disease and fail for another. Oral semaglutide, for example, recently failed to demonstrate efficacy in Phase III trials for early-stage Alzheimer's disease despite being previously FDA approved for the treatment of type 2 diabetes mellitus in adults.[6,7] A useful forecasting approach for our problem must be able to distinguish between these outcomes. Furthermore, because this decision is made before human trials begin, a model intended to support it cannot rely on protocol-derived features, trial outcomes, or other evidence generated after clinical development has started.

Prior drug outcome forecasting work typically divides into three main approaches: (1) drug-level outcomes such as clinical toxicity or eventual FDA approval[8-12], (2) ranking diseases based on historical drug-disease relationships[13-15], or (3) predicting whether a particular clinical trial will succeed.[16-25] Drug-level models use combinations of molecular, physicochemical, clinical-trial, patient, and language-model features, assigning one score per molecule without distinguishing disease-specific outcomes.[8-12] These models can forecast whether a drug may be approved for any disease, but typically do not discriminate between different disease-level outcomes. Although disease-resolved studies remain less common, a few studies have modeled outcomes at the drug-disease level. Lo et al. use Phase II or III information to predict eventual approval,[26] while inClinico predicts drug-disease transition from Phase II to Phase III.[17] Crucially, because both rely on information generated during clinical development, they do not tackle our pre-clinical, trial-agnostic outcome forecasting question.

Disease-ranking approaches better align with our prediction unit, but generally optimize based on therapeutic association and relevance rather than eventual FDA approval. Methods such as PREDICT infer therapeutic drug-disease associations,[15] while INSPIRE ranks candidate diseases for a therapeutic mechanism using real-world patient-data embeddings and evaluates those rankings against subsequent approvals; its score reflects predicted therapeutic relevance rather than the probability of eventual approval across drug-disease programs.[14] RepurposeDrugs most closely overlaps with our task: it uses molecular descriptors and disease features to score drug-disease pairs by approval likelihood, but its published evaluation does not provide a reproducible temporal benchmark of disease-level FDA outcomes or directly evaluate whether the scores correctly discriminate between disease outcomes for a given drug.[13]

Trial-outcome approaches span a wide range of input modes but generally operate on registered trials and trial-level endpoints rather than full end-to-end approval.[16-25] TrialBench and the Clinical Trial Outcome benchmark, for example, expand trial-level prediction and outcome labeling, but do not define final approval at the drug-disease level.[24,25] More broadly, trial-outcome models use protocol attributes including eligibility criteria, phase, and sponsor as inputs to evaluate programs that have already entered development rather than informing the earlier pre-clinical choice of which programs to pursue.[16-25,27,28]

A separate line of research focuses on historical likelihood-of-approval (LOA) rates instead of outcome forecasting. This work examines historical trends, aggregating drug approval numbers by phase, disease area, modality, sponsor type, time window, and, in some studies, target and action.[4,5,29-33] The most detailed LOA analyses resolve drug programs by disease and inform parts of the methodology in this work,[5,32,33] but they remain population-level analyses rather than forecasts for individual drug-disease pairs. Table 1 summarizes this detailed landscape. These approaches leave unresolved whether future FDA approvals can be forecasted for individual drug-disease programs before clinical trials begin and whether those forecasts provide useful signal for choosing among candidate diseases for the same drug.

We test two central claims. First, FDA approval for an individual drug-disease program is partially predictable before human trials begin using public, trial-agnostic information. Second, drug-disease approval forecasting provides useful signal for prioritizing between candidate diseases for a chosen drug. We present an approach to drug approval forecasting from publicly available data under two constraints: (1) the prediction is made for a particular combination of drug molecule and disease, and (2) it is made before clinical trials begin (Fig. 1).

This work can be divided into four main contributions. First, because existing public data generally provides approval outcomes at the drug level rather than the drug-disease level, our task requires us to build a dataset which supports our decision task. We present a pipeline which reconstructs drug-disease candidates from ClinicalTrials.gov via AACT, assigns disease-level approval labels extracted from FDA label information, and attaches molecular, disease, ADMET, target, and pathway features pulled from public sources (Fig. 2). Second, we evaluate post-2019 approval forecasting using a model trained on molecule and disease features alone as the cleanest pretrial benchmark. We further expand this analysis to our full feature set and compare performance to adapted models from adjacent tasks including ChemAP and HINT.[11,22] Third, we analyze which feature groups carry predictive signal and evaluate performance for drugs seen and unseen during training. Fourth, we test whether the same approval forecasts can support disease prioritization for a single drug by evaluating within-drug disease rankings and examining the outcomes of the highest scoring candidates. We supplement these disease comparisons with case studies and a public web tool for forecasting drug-disease approval probabilities from only a drug name and target disease.

## Results

### *A public drug-disease cohort defines the approval-forecasting task*

While existing benchmarks such as HINT and TrialBench support trial-outcome modeling,[22,24] they lack disease-level final-approval labels. That is, a drug indicated for disease 1 but not disease 2 is treated as a single sample, despite having disease-level differences. In order to model these nuanced differences, we define our prediction unit as a drug-disease pair rather than an independent drug, trial, or disease-area. Our endpoint is disease-level FDA approval status for the given drug up to the cutoff date (2026-05-07). We further restrict ourselves to trial-agnostic predictions, targeting the decision point before human trials start and before eligibility criteria or sponsorship information are known (Fig. 1b).

We built our cohort from interventional drug trials in ClinicalTrials.gov via AACT (Fig. 2a). Trial records were normalized into drug-disease pairs, approval status was extracted from FDA label data via an automated LLM-based pipeline, pairs were filtered to those with both a molecular representation and an ICD-10 disease code, and the cohort was split temporally, with post-2019 pairs held out as the test set. Molecular, disease, ADMET, target, and pathway features were then attached for modeling. In order to generate molecular fingerprints, ADMET predictions, and target and pathway annotations, we required a SMILES string for every drug and filtered out those without a match. We note that this biases the cohort toward small molecules with known SMILES and better documented molecular structures which in turn inflates the overall dataset approval rate (17.5% approved) relative to previous LOA analyses (e.g., 10.4% from Phase I in Hay et al.).[5] A standardized disease code was likewise required for the drug-disease prediction task. These filters improve consistency for the present benchmark but limit generalization to records without standardized disease codes and to complex modalities not well represented by SMILES.

The final dataset contains 14,134 resolved drug-disease pairs: 2,469 approved and 11,665 unapproved at the study cutoff. The held-out test set contains 2,637 pairs, including 624 approvals. The cohort spans 3,238 unique drugs and 4,190 unique diseases, with an average of 4.37 diseases tested clinically per drug. This repeated-disease structure is central to the task: because many drugs appear across multiple diseases, a useful model must distinguish disease-specific outcomes rather than assign a single fixed approval tendency to each molecule.

To assess the reliability of the disease-level labels, we manually reviewed a stratified sample of 100 pipeline decisions, including 50 labeled approved and 50 labeled not approved. Manual adjudication confirmed 48 of 50 approved calls (PPV 96.0%) and 46 of 50 not-approved calls (NPV 92.0%; Supplementary Methods S7.3).

***Preclinical data predicts drug-disease approval outcomes***

For comparison with existing open-source models, we adapted models trained on adjacent tasks to our dataset as baselines. We examined two, HINT and ChemAP, and compared them to a simple tree-based model trained directly on our dataset (Fig. 3).[11,22] These models were originally trained for clinical trial success and molecule-level approval outcomes respectively and therefore form a useful benchmark for evaluating whether our drug-disease task is meaningfully distinct. The key result is that these adjacent-task models did not translate to our drug-disease level pre-Phase I task and were both outperformed by our model trained directly for end-to-end, drug-disease approval forecasting. Values are reported as point estimates with 95% confidence intervals in brackets. On the held-out drug-disease test set, our model trained on our full feature set achieved ROC-AUC 0.77 [0.75, 0.80] and PR-AUC 0.56 [0.51, 0.61] (Fig. 3; Table 2). A molecule-only model achieved ROC-AUC 0.72 [0.69, 0.75] and PR-AUC 0.42 [0.37, 0.47]. Adding disease identity improved performance to ROC-AUC 0.77 [0.75, 0.80] and PR-AUC 0.55 [0.49, 0.61], validating that disease context is in fact a meaningful differentiator between drug-disease outcomes. The low overall PR-AUC values can be partly attributed to the low baseline approval rates in our cohort.

Because unresolved programs are excluded, more recent test cases are biased toward drug development programs that resolved quickly. We therefore repeated the evaluation on a subset of our test data including only the 1,486 programs that began in 2020-2021. This subset had a reduced 14.1% approval rate. ROC-AUC was stable or higher for every evaluated model: 0.79 [0.75, 0.82] for the full model, 0.80 [0.76, 0.83] for molecule + disease, 0.73 [0.69, 0.77] for molecule only, 0.62 [0.58, 0.67] for adapted HINT, and 0.46 [0.42, 0.51] for ChemAP. Full-model PR-AUC decreased from 0.56 to 0.45 as a product of the lower approval prevalence; intermediate start-year restrictions showed the same pattern (Supplementary Methods S10).

Two baseline models, originally designed for different but related tasks, were not able to match this performance. First, we adapted the published HINT architecture which was originally trained for predicting trial-level outcomes[22] to our drug-disease approval task. We removed the eligibility-criteria input and retrained the remaining model on our disease-level approval labels. The retrained model achieved ROC-AUC 0.62 and PR-AUC 0.33 on our end-to-end, drug-disease task. This should not be read as a failure of HINT as HINT was not originally designed for this task and part of HINT's strengths is in leveraging trial eligibility criteria which, as previously mentioned, was explicitly excluded. Rather, the result shows that a model designed for same-trial outcome prediction does not directly solve final drug-disease approval forecasting. To further validate that these results were not an issue with our implementation of HINT, we trained the same eligibility-stripped version of the model as well as our tree-based model on the TOP benchmark used in the HINT study and found that both models performed similarly across phases (Supplementary Table S4).

To test whether drug-level predictions transfer to the drug-disease task, we applied ChemAP, which predicts FDA approval from molecular structure alone, without retraining.[11] ChemAP achieved ROC-AUC 0.46 [0.43, 0.50] and PR-AUC 0.22 [0.19, 0.25]. The below-chance ROC-AUC is consistent with a task mismatch: this benchmark labels each drug-disease pair separately, and a molecule approved for one disease may be unapproved for others. Assigning a single high score to every disease for an approved molecule can therefore become anticorrelated with disease-level approval labels.

To confirm that this result reflects a task mismatch rather than a ChemAP-specific modeling issue, we also evaluated a hypothetical model with perfect drug-level labels. We assumed the hypothetical model is perfect at determining if a drug will reach FDA approval, but has no concept of disease-level approval outcomes. Under these assumptions, the hypothetical model still underperformed our own task-specific model (ROC-AUC 0.71; PR-AUC 0.35). These comparisons reinforce the central motivation for disease-specific forecasting. Drug-level approval forecasts do not capture the full picture of drug development outcomes and are clearly outperformed by disease-aware models when it comes to forecasting outcomes at the drug-disease level.

***Molecular and disease features carry most of the predictive signal***

To understand which feature groups drive our model predictions, we trained models on different feature subsets (Fig. 4; Table 2). As single feature groups, molecule and ADMET were the strongest signals (ROC-AUC 0.72 and 0.71, respectively). Disease, target, and pathway features, when used in isolation, carried weaker but non-zero signal (ROC-AUC 0.66, 0.63, and 0.59 respectively). Combining feature groups improved on any single group: molecule + disease reached ROC-AUC 0.77, adding target features gave 0.77, and the full molecule + disease + target + pathway + ADMET model reached 0.77. The full model did not improve on molecule + disease under a paired drug-cluster bootstrap analysis. These results indicate that public target and pathway annotations, as defined in our feature set, carry meaningful information about approval but do not add detectable improvement once molecular and disease features are present. Because molecular structure and disease identity are directly available before trials, the molecule + disease model provides the most conservative test of the pretrial forecasting claim.

We also stratified test performance by whether a drug had appeared in training (Fig. 4c, f). Since previously approved drugs can and often are repurposed for new diseases, we did not explicitly exclude these from our full test set. The full model achieved ROC-AUC 0.78 for seen drugs and 0.71 for unseen drugs. Performance therefore benefits from related drugs in training but does not collapse on unseen drugs.

### *Approval forecasts inform disease prioritization*

Beyond evaluating individual drug-disease programs, we asked whether the same forecasts could help choose among candidate diseases for a given drug. To evaluate this decision directly, we ranked candidate diseases within each drug and performed a "top-1" analysis by selecting the highest-scoring disease for each drug. Across 212 drugs with at least one known approved and one known unapproved disease, the full model selected an approved disease in 61.3% [54.7%, 67.9%] of cases, compared with 50.9% [44.3%, 58.0%] using disease-only predictions and 40.7% [38.1%, 43.1%] under random selection (Table 3).

Across all 1,721 approved-unapproved disease pairs for these drugs, the full model achieved a within-drug AUC of 0.71 [0.66, 0.76]. Disease-only and molecule + disease models achieved within-drug AUCs of 0.67 [0.63, 0.71] and 0.71 [0.65, 0.76], respectively. Our baseline molecule-only predictions yield an AUC of 0.50 due to disease not being represented in the inputs. Disease-only performance shows that disease-level approval trends contribute substantial ranking signal, but still fall short of models that also include the drug when evaluating top-1 disease-selection decisions.

Selected case studies can illustrate how this behavior emerges. For tacrolimus, approved transplant-related diseases including transplant rejection (0.50) and kidney transplant failure (0.44) ranked far above autoimmune hepatitis (0.04), an unapproved disease with ongoing clinical interest. Tacrolimus is FDA-approved to prevent rejection after kidney, liver, heart, and lung transplantation.[34] While treatment of autoimmune hepatitis has been explored off-label, it remains under clinical study.[35,36] Another example, salbutamol, shows the model gave much higher approval probabilities for an approved disease, asthma (0.63), than a currently unapproved disease renal colic (0.20).[37]

The model also differentiated outcomes within the same molecule even when its absolute probabilities may be over or under inflated. Pomalidomide received a probability of 0.69 for multiple myeloma and 0.59 for primary central nervous system lymphoma. The current Pomalyst label includes multiple myeloma and Kaposi sarcoma, but not primary central nervous system lymphoma.[38] A Phase I study of pomalidomide plus dexamethasone in relapsed or refractory primary central nervous system or vitreoretinal lymphoma nevertheless reported clinical activity.[39] At the time of writing, however, the drug is not approved for treatment of this disease. Despite the relatively high approval probability for this unapproved disease, the model still preserves accurate within-drug ranking.

One example illustrates the current limitations of our approach. For treprostinil, the model ranked several unapproved diseases above its approved uses: pulmonary arterial hypertension and pulmonary hypertension associated with interstitial lung disease.[40] Inhaled treprostinil has demonstrated benefit in the latter population,[41] which can include patients with idiopathic pulmonary fibrosis. However, treprostinil is not currently approved for idiopathic pulmonary fibrosis itself.[40] More recent phase III trials of inhaled treprostinil in idiopathic pulmonary fibrosis reported positive results, illustrating that a disease classified as unapproved at the dataset cutoff may later resume development or produce favorable evidence.[42] These errors occur between closely related pulmonary diseases, suggesting that disease-specific distinctions remain challenging when candidate diseases share substantial biological similarity.

Taken together, these examples show that the model is capable of prioritizing diseases for an individual drug rather than relying on a generic drug-level success signal. At the same time, the failure cases demonstrate situations where the current feature representation provides insufficient resolution to distinguish between closely related diseases, highlighting an important direction for future work.

***A public web tool exposes our trial-agnostic approval forecasting***

To enable broader use of the model for vetting drug-disease candidates, we provide a publicly available web tool that exposes the model and its predictions at https://huggingface.co/spaces/samirkt/drug-disease-approval-forecasting. A user can enter a drug name and disease, which are resolved to a SMILES representation and an ICD-10 code, respectively, or supply a SMILES string and ICD-10 code directly. For pairs with a known outcome in the model-development data, the tool reports that observed outcome. Otherwise, it returns a predicted approval probability for the drug-disease pair. The web tool is not a final clinical, regulatory, or investment decision system. Its purpose is to make the model's insights usable, to allow comparison of disease hypotheses for the same molecule, and to provide additional signal on promising development paths. The web tool, at the time of publishing, serves the molecule + disease model to balance ease of use with performance.

## Discussion

Our results support two uses of preclinical approval forecasting. First, approval outcomes for drug-disease programs in our dataset were predictable on temporally held-out test data based on our full feature set (ROC-AUC 0.77; PR-AUC 0.56), supporting their potential use in early program evaluation. The molecule + disease model (ROC-AUC 0.77; PR-AUC 0.55), meanwhile, matched the full model (0.77; 0.56) while providing a clearer pre-clinical translational option by only requiring a SMILES string and ICD-10 code. Second, those forecasts improved disease selection: the model's top-ranked disease was approved in 61.3% of eligible drugs, compared with 50.9% using a disease-only model and 40.7% under random selection. Performance also exceeded ChemAP applied out of the box and a HINT-style model adapted to make predictions without trial eligibility criteria, demonstrating that adjacent drug outcome forecasting tasks do not necessarily generalize to our drug-disease level decision.[11,22]

The prediction unit matters. Drug-level models cannot distinguish diseases for the same molecule, and trial-outcome models may depend on information unavailable before development begins. The drug-disease formulation instead matches an early program-advancement decision: should this molecule be tested in human clinical trials for this disease? Comparing those predictions across diseases additionally enables disease selection.

Disease information alone was sufficient for a model to learn to rank diseases reasonably well, achieving a within-drug AUC of 0.67, somewhat below the molecule + disease model (0.71). However, both the molecule + disease model and the full feature model were better at selecting the top-ranked disease. This suggests that some diseases are generally more likely to succeed than others, but information about the drug still improves the choice most directly aligned with disease selection. This is the clearest evidence for the value of drug information alongside disease identity for disease selection.

The feature analyses suggest that most of the available signal comes from molecular structure, ADMET predictions, and disease context. This is consistent with prior work linking molecular complexity and three-dimensional descriptors to clinical progression, and chemical predictive models to compound-quality optimization.[43-45] Target and pathway features carry signal in isolation, but their incremental value is modest in the current representation. This should not be taken as evidence that biological mechanism is unimportant to approval: target-disease studies have repeatedly associated human genetic or integrated evidence with clinical success.[46-48] A more cautious interpretation is that the public target and pathway annotations used in this study are not sufficiently complete or independent to add performance improvements beyond what is achievable using molecular and disease features.

RepurposeDrugs previously demonstrated approval-likelihood scoring of drug-disease pairs. Relative to that work, the present evaluation adds a public disease-level FDA outcome benchmark, temporal held-out evaluation for prospective analysis, and direct testing of same-drug disease selection. The distinction is therefore in the evaluation design, outcome benchmark, and model performance rather than in introducing drug-disease approval scoring itself.

Several limitations constrain these conclusions. The cohort is retrospective and depends on the coverage of ClinicalTrials.gov, AACT, FDA label data, DrugBank, ChEMBL, Open Targets, Reactome, ICD-10, and ADMET-AI. The cohort is reconstructed from registered clinical programs and is therefore not a true random sample of possible drug-disease hypotheses. ClinicalTrials.gov records have known completeness and standardization limitations that complicate secondary reuse and can propagate into program reconstruction.[49,50] The ICD-10 requirement excludes many constructed candidate records, and the SMILES requirement limits coverage of biologics, mixtures, and other complex modalities. Disease-level labels rely on automated FDA-label matching and language-model information extraction from unstructured labels, so some residual label noise is expected. The seen-versus-unseen analysis suggests generalization beyond exact-molecule memorization, but unseen-drug performance is lower and the unseen subset is small. ADMET, target, and pathway features were derived from contemporary resources and represent information that could plausibly be available before clinical trials. Because these resources evolve over time, exact historical availability cannot be reconstructed for every program. The unapproved class is heterogeneous and includes programs that were discontinued, inactive, incompletely developed, or simply not approved by the study cutoff. These states should therefore not all be interpreted as demonstrated efficacy failures.

Despite these limitations, the study provides a public disease-level benchmark and forecasting model for evaluating drug-disease programs before clinical trials begin. Together, the results support preclinical approval forecasting as a tool for evaluating whether drug-disease programs should advance and for prioritizing among candidate diseases. Future work should extend coverage beyond SMILES-representable molecules, explore alternative feature representations and embeddings, and test richer target and pathway features.

## Methods

Full procedural detail for the dataset-construction pipeline and modeling layer is provided in the Supplementary Methods. The accompanying repository contains the released dataset, trained models, and analysis code (see Data availability). The summary below describes the final analysis corresponding to the main-text figures and tables.

### *Clinical-trial ingestion and drug-disease construction*

We used the AACT relational database of ClinicalTrials.gov[51,52] to ingest interventional drug trials and expanded each record into study × drug × condition tuples. Placebo rows, healthy-volunteer rows, and non-drug interventions were removed. Drug names and molecular identifiers were normalized using DrugBank and ChEMBL as reference resources, including available synonyms, with fallback normalization of free-text intervention names.[53,54] Conditions were normalized using MeSH terms and normalized free-text condition names, with ICD-10-CM codes assigned subsequently.[55] Normalized drug and disease names were used to match related clinical trials, adopting the drug-disease development-path unit used by Hay et al.[5]

***Approval labeling***

Because openFDA exposes approved diseases as unstructured label text, candidate outcomes were assigned by matching drug-disease pairs to FDA label information.[56] For each drug, we retrieved approved disease text from a local FDA-label mirror.[56] Label text was segmented into disease strings, and a locally hosted Qwen 2.5-7B-Instruct model[57] was used to determine whether each candidate drug-disease pair was covered by the approved label. The full adjudication prompt is provided in Supplementary Note 1. The broader strategy is related to recent automated trial-outcome labeling work, but our endpoint is resolved at the drug-disease program level rather than the individual-trial level.[25] Candidates were labeled positive if the drug was approved for the matched disease, and negative if the disease was not approved at the study cutoff according to the resolved trial trajectory. Recently ongoing, ambiguous, or otherwise unresolved candidates were excluded from the main supervised dataset. The label-segmentation, language-model adjudication, and manual-validation procedures are described in Supplementary Methods S7.1-S7.3.

***Filtering and temporal split***

The final cohort was restricted to resolved drug-disease pairs with both an available molecular structure and an ICD-10 disease code, and to post-2009 programs based on FDA-label availability. The cohort contains 14,134 drug-disease pairs (2,469 approved; 11,665 unapproved at the study cutoff). Evaluation used a temporal split designed to mimic prospective deployment: programs starting in or before 2019 were used for model development, and programs starting after 2019 were held out for testing. The held-out test set contains 2,637 pairs, including 624 approvals.

### *Feature representation*

Features were grouped into five categories: molecule, disease, ADMET, target, and pathway. Molecular structure was represented from SMILES using RDKit-derived fingerprints, including ECFP4 and MACCS keys.[58] Disease context was represented using ICD-10-CM codes accessed through the U.S. National Library of Medicine's Clinical Tables API,[59] with the underlying code set maintained by the CDC's National Center for Health Statistics.[60] ADMET features were generated from molecular structure using ADMET-AI.[61] Target annotations were derived from ChEMBL and Open Targets, including available target names and target-level evidence.[54,62,63] Pathway annotations were derived from Open Targets and Reactome.[63,64] Molecular structure and disease identity are directly available at the pretrial decision point. ADMET, target, and pathway features come from contemporary resources and represent additional information that could plausibly be available before trials. String and categorical features were one-hot or multi-hot encoded. To reduce risk of overparameterization in our dataset, high-dimensional feature groups were compressed: dense groups such as molecular fingerprints with PCA, and sparse groups such as the multi-hot ICD-10 encodings with truncated SVD. In all cases, high-dimensional feature representations were reduced to 50-element vectors.

### *Models and training*

The primary model was a gradient-boosted decision-tree classifier implemented with XGBoost.[65] Hyperparameters were fixed a priori and held constant across experiments so that comparisons reflected feature sets and task framing rather than per-model tuning. Class imbalance was handled by up-weighting positive examples in proportion to the negative-to-positive ratio in the development set. A single stratified 10% slice of the training set was held out. Encoders, dimensionality-reduction transforms, and the final classifier were all fit on the remaining 90%. The held-out slice was used solely for gradient-boosting early stopping. For the molecule + disease model served by the public web tool, probabilities were calibrated with Platt scaling.[66] fit on five-fold out-of-fold predictions over the training set (Supplementary S13). No model selection was performed on the held-out slice, no component was refit on the complete training set, and the post-2019 test set was untouched by all fitting. Main discrimination comparisons use raw model scores. Because recent test programs are right-censored (only fast-resolving programs appear, and these are predominantly approvals), calibration was assessed on the 1,486 test programs starting 2020-2021: calibration reduced the Brier score from 0.12 to 0.10 and expected calibration error from 0.14 to 0.02 (Supplementary Fig. S1).

### *Transferred baselines*

Two adjacent-task baselines tested whether existing models solve the disease-specific approval problem. HINT was adapted to the trial-agnostic drug-disease task by removing the eligibility-criteria input and using only the available drug and disease inputs. To confirm that the comparison did not depend on a broken HINT implementation, we also evaluated HINT and a simple XGBoost comparator on the HINT-native Phase 1, Phase 2, and Phase 3 benchmarks. Those results are reported in Supplementary Table S4. ChemAP was applied out of the box as a drug-level approval model. Because it does not model disease-specific labels, each drug-level score was compared against the corresponding drug-disease label in our test set.

### *Disease-selection evaluation*

Disease-selection performance was evaluated among drugs with at least one approved and one unapproved disease in the held-out test set. Within-drug AUC was calculated across approved-unapproved disease pairs while holding drug identity fixed. An additional top-1 analysis ranked candidate diseases by predicted approval probability for each drug, and the highest-scoring disease was counted as a hit if it was approved. We compared the full model with disease-only predictions and random selection among each drug's observed candidate diseases. Confidence intervals and paired differences were estimated by bootstrap resampling at the drug level.

### *Evaluation and uncertainty*

Models were evaluated using ROC-AUC and PR-AUC as primary discrimination metrics. F1 scores and threshold-dependent summaries are reported in Supplementary Table S3. ROC-AUC and PR-AUC confidence intervals for all model comparisons were computed by bootstrap resampling drugs with replacement and retaining all associated test-set diseases. Paired model differences used the same drug-cluster bootstrap replicates. F1 confidence intervals were computed by candidate-level bootstrap resampling. The seen-versus-unseen analysis stratified held-out predictions by whether the drug appeared in the training set. Case studies were selected as qualitative examples of disease-level ranking and were not used as primary evidence for model performance.

## Funding

This work was supported by the USDA-NIFA AI Institute for Next Generation Food Systems (AIFS), USDA-NIFA award number 2020-67021-32855.

## Conflict of interest

The authors declare no competing interests.

## Data availability

The drug-disease cohort, approval labels, trained models and experiment code are available at https://github.com/IBPA/drug-approval-forecast/tree/main and archived at DOI: 10.5281/zenodo.22967987. The web tool is available at https://huggingface.co/spaces/samirkt/drug-disease-approval-forecasting. The source datasets and resources used in this study, together with their versions or retrieval dates and applicable access conditions, are listed in Supplementary Methods S2.

## Tables

**Table 1.** Related work in drug approval prediction, disease prioritization, and clinical-development outcome prediction. Prior studies predict whether drugs succeed, rank possible diseases by therapeutic properties, or forecast outcomes after clinical development begins. The present study adds temporal evaluation of disease-level approval forecasts and direct within-drug disease-selection testing.

| Study | Prediction Task | Inputs |
|---|---|---|
| Hay et al.[5] | Historical LOA and phase-transition rates by phase and disease area | Industry-curated development histories; clinical phase; disease area; lead/non-lead disease; regulatory status |
| Wong et al.[4] | Clinical success-rate estimation from curated trial/program histories | Citeline (Trialtrove/Pharmaprojects) trial and approval histories; disease type; phase; sponsor type; biomarker use; lead disease; time |
| BIO / Informa / QLS[30] | Industry-standard LOA rates by phase, disease, and modality | Industry-curated clinical histories; disease; modality; phase; regulatory and program factors |
| Dowden & Munro[31] | Therapeutic-area and pipeline productivity trends | Therapeutic area; company pipeline trends; approval and development success trends |
| Zhou et al. / ClinSR[32] | Dynamic clinical success-rate reconstruction | ClinicalTrials.gov histories; FDA approvals; drug/disease normalization; modality; disease class; time |
| Yamaguchi et al.[33] | Approval success rates stratified by target, action, modality, and application | Candidate target; pharmacological action; modality; application type; development and approval status |
| Lo et al.[26] | Later-stage approval prediction for drug-disease programs | Proprietary Pharmaprojects/Trialtrove data; compound attributes; trial characteristics; sponsor history; completed Phase II/III status |
| Gottlieb et al. / PREDICT[15] | Drug-disease prediction for approved and novel molecules | Drug-drug and disease-disease similarity measures; known therapeutic associations |
| Ianevski et al. / RepurposeDrugs[13] | Non-prospective drug–disease approval prediction | Molecular fingerprints and physicochemical descriptors; disease identity and class; approved and terminated diseases |
| Eckhoff et al. / INSPIRE[14] | Early disease ranking validated against subsequent approvals | Real-world patient-data embeddings; reference diseases; mechanism-of-action-specific disease ranking |
| John et al.[10] | Drug-level failure prediction during clinical development | Molecular descriptors and fingerprints |
| ChemAP / Cho et al.[11] | Drug-level approval prediction | Chemical structure and fragments at inference; multimodal teacher trained with physicochemical, patent, and trial features |
| Ghaffarzadeh-Esfahani et al. / DrugReasoner[12] | Interpretable drug-level approval prediction | Molecular descriptors; MolFormer-based similarity retrieval; reasoning-augmented language model |
| DiMasi et al. / ANDI[27] | Oncology approval algorithm for a drug or lead disease | Four-factor oncology Phase II decision variables |
| Beinse et al. / RESOLVED2[28] | Time to FDA approval after Phase I oncology trials | Phase I abstracts and DrugBank pharmacology; penalized Cox model |
| Aparício et al. / MMCTOP[16] | Multimodal clinical-trial-outcome prediction | Molecular structure, protocol metadata, eligibility text, and disease ontologies |
| Aliper et al. / inClinico[17] | Drug-disease Phase II-to-III transition prediction | Target choice, omics, text, trial design, small-molecule properties, and proprietary ensemble inputs |
| Wang et al. / SPOT[18] | Trial success/failure prediction with similar-trial histories | Diseases, treatments, eligibility, title, outcomes, descriptions, and historical trial sequences |
| Lu et al.[19] | Confidence-aware trial-outcome prediction | HINT/TOP-style drug, disease, and protocol inputs with selective-classification uncertainty |
| Gao et al. / LINT[20] | Biologics trial-outcome prediction | Free-text trial descriptions |
| Reinisch et al. / CTP-LLM[21] | Large-language-model phase-transition prediction | Original protocol text across phases |
| Fu et al. / HINT[22] | Clinical-trial-outcome prediction | Molecule, disease, eligibility criteria, ADMET/PK, and historical-trial inputs at trial start |
| Zheng et al. / LIFTED[23] | Multimodal clinical-trial-outcome prediction | Textualized molecular, disease, eligibility, and protocol modalities; mixture of experts |
| Chen et al. / TrialBench[24] | Trial-outcome benchmark datasets | SMILES, eligibility text, MeSH/ICD-10, and categorical and numerical trial data |
| Gao et al. / CTO[25] | Automated trial-outcome labeling and benchmark construction | ClinicalTrials.gov, publications, cross-phase linkage, regulatory and other outcome signals, and language-model labeling |

**Table 2.** Main held-out drug-disease performance. All metrics are reported on the temporally held-out post-2019 test set of 2,637 drug-disease pairs (624 approved). ChemAP was applied out of the box as a drug-level approval model. HINT was adapted by removing the eligibility-criteria input. All 95% confidence intervals were computed by drug-cluster bootstrap resampling.

| Model / feature set | Task origin or feature class | ROC-AUC (95% CI) | PR-AUC (95% CI) |
|---|---|---|---|
| ChemAP | Drug-level approval model applied out of the box | 0.462 [0.425, 0.497] | 0.218 [0.191, 0.252] |
| HINT adapted | Trial-outcome model adapted by removing eligibility input | 0.619 [0.591, 0.647] | 0.334 [0.294, 0.379] |
| Molecule only | Single feature group | 0.721 [0.691, 0.747] | 0.419 [0.367, 0.471] |
| ADMET only | Single feature group | 0.714 [0.685, 0.741] | 0.402 [0.354, 0.454] |
| Disease only | Single feature group | 0.656 [0.628, 0.685] | 0.359 [0.319, 0.407] |
| Target only | Single feature group | 0.626 [0.593, 0.657] | 0.348 [0.307, 0.398] |
| Pathway only | Single feature group | 0.593 [0.563, 0.625] | 0.326 [0.285, 0.368] |
| Molecule + disease | Task-specific simple model | 0.773 [0.748, 0.797] | 0.550 [0.494, 0.606] |
| Mol + disease + target | Task-specific cumulative model | 0.769 [0.742, 0.793] | 0.548 [0.494, 0.600] |
| Full model | Molecule + disease + target + pathway + ADMET | **0.771 [0.745, 0.796]** | **0.559 [0.506, 0.611]** |

**Table 3.** Within-drug disease-selection performance. Within-drug AUC compares approved and unapproved diseases while holding the drug fixed. Top-1 hit rate reports the fraction of eligible drugs for which the highest-ranked observed candidate disease was approved. Confidence intervals were estimated by drug-level bootstrap resampling.

| Strategy | Within-drug AUC (95% CI) | Top-1 hit rate (95% CI) |
|---|---|---|
| Disease only | 0.672 [0.628, 0.711] | 0.509 [0.443, 0.580] |
| Molecule + disease | 0.711 [0.654, 0.764] | 0.599 [0.538, 0.665] |
| Full model | 0.709 [0.656, 0.758] | 0.613 [0.547, 0.679] |
| Random | - | 0.407 [0.381, 0.431] |

## Figure Captions

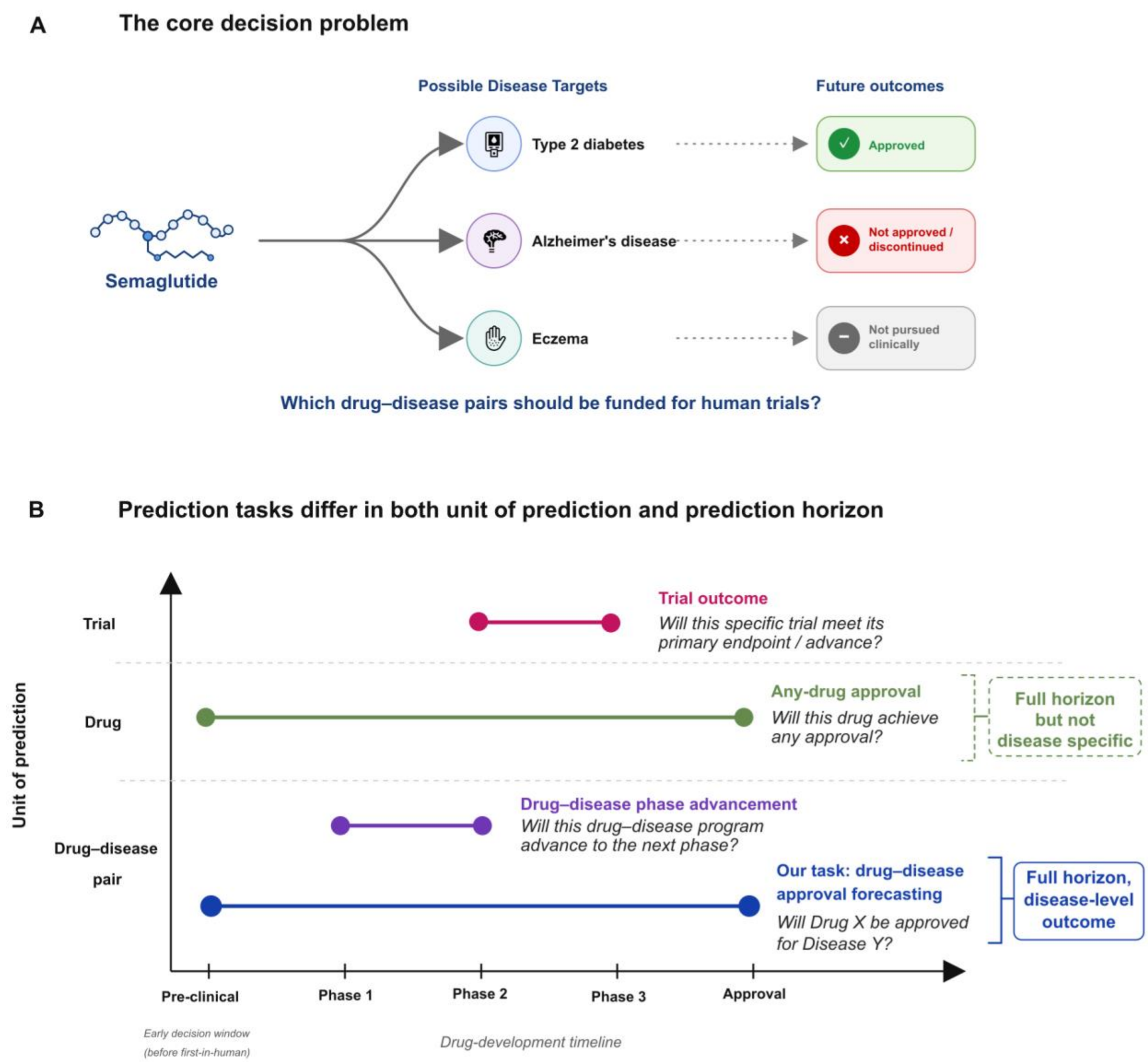

**Figure 1.** Framing the preclinical drug-disease approval forecasting problem. **(a)** A single drug may face different eventual approval outcomes for different candidate diseases. The early prioritization of which drug-disease pairs should advance to human testing is critical to minimizing expected costs. **(b)** Drug development forecasting tasks differ in both prediction unit

and prediction horizon. Trial-outcome models evaluate individual trials, forecasting if a trial will meet its primary endpoints. Drug-level models forecast whether a molecule achieves any approval for any disease. Some phase-advancement models evaluate a drug-disease pair over a segment of the clinical development lifecycle. Our task forecasts eventual FDA approval for a specific drug-disease pair from the preclinical decision window. Solutions to this task can support early program assessment and disease prioritization before trial-specific information is available.

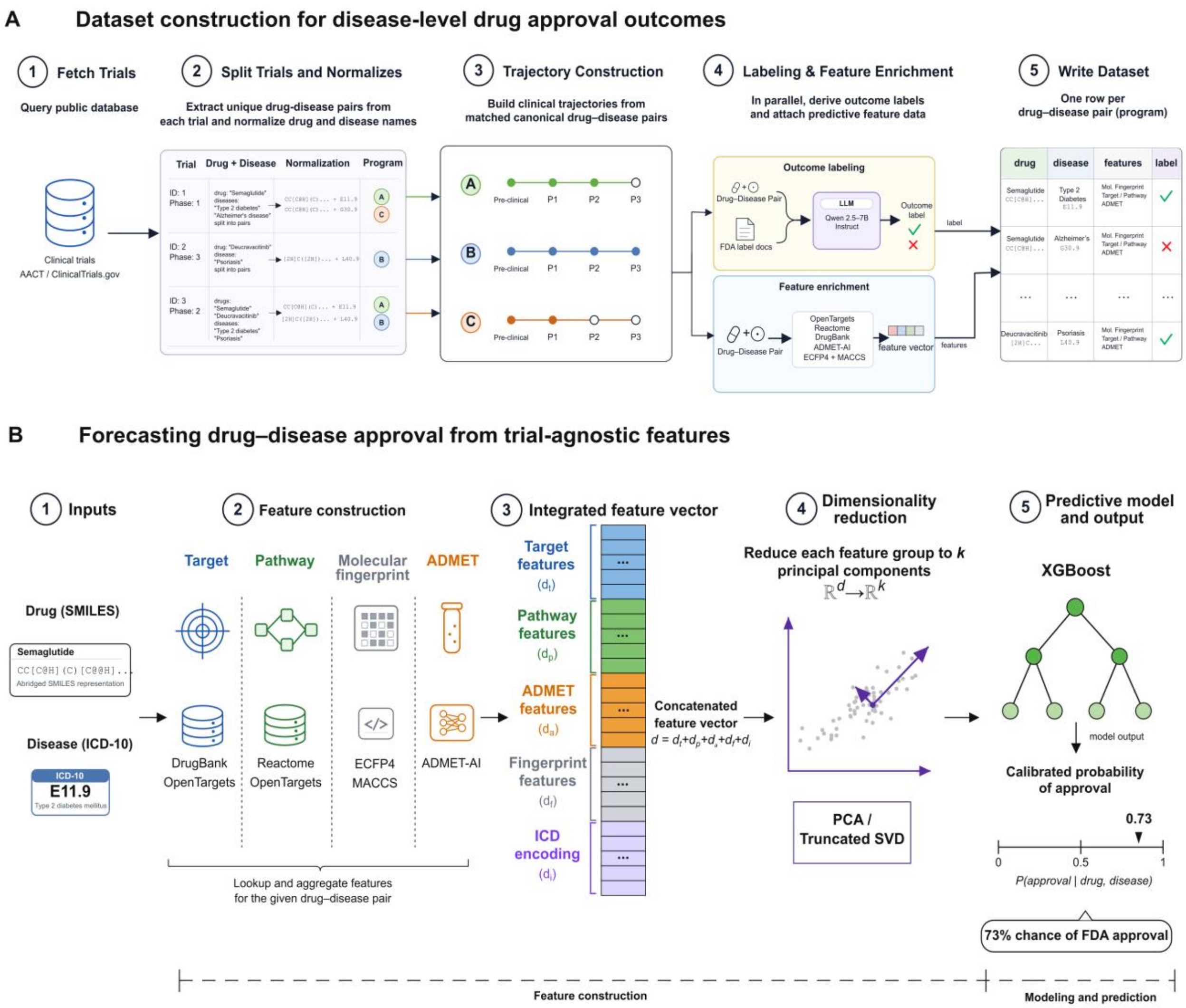

**Figure 2.** Dataset construction and model inference. (**a**) Interventional trials from ClinicalTrials.gov are normalized by drug and disease name and linked across phases to identify the minimum set of drug-disease pairs in our data. Drug-disease pairs are assigned disease-specific approval labels by extracting diseases from unstructured FDA label documentation using Qwen 2.5-7B Instruct. Each sample is enriched with additional feature data from open source models and databases and written as one drug-disease row in our database. (**b**) At inference, a drug (represented as SMILES) and disease (ICD-10 code) are used to gather and construct unique feature groups for the drug-disease pair: target, pathway, molecular-fingerprint, and ADMET. Each feature group independently goes through dimensionality reduction via PCA or truncated SVD, and the resulting concatenated vector is passed to an XGBoost classifier. Feature groups are manually preserved or excluded for different model experiments. Held-out calibration is evaluated in Supplementary Fig. S1.

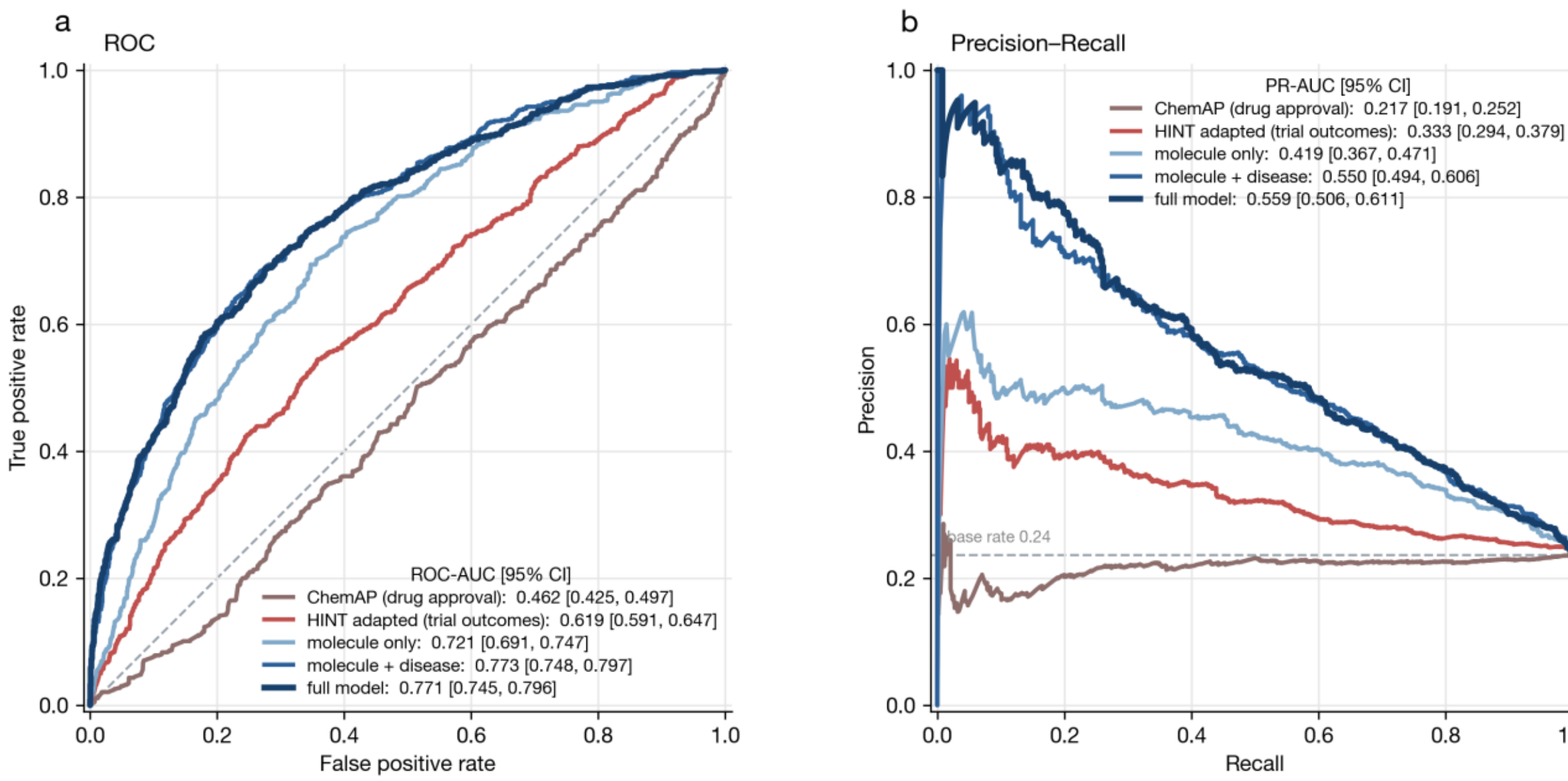

**Figure 3.** Task-specific drug-disease forecasting versus adjacent-task baselines. (**a**) ROC curves and (**b**) precision-recall curves on the held-out post-2019 drug-disease test set. Evaluated models include ChemAP applied out of the box, eligibility-stripped HINT, and our simple XGBoost model trained with several different feature group combinations. Our task-specific models consistently outperform the transferred baselines. AUC values and 95% bootstrap confidence intervals are reported in the legends and in Table 2.

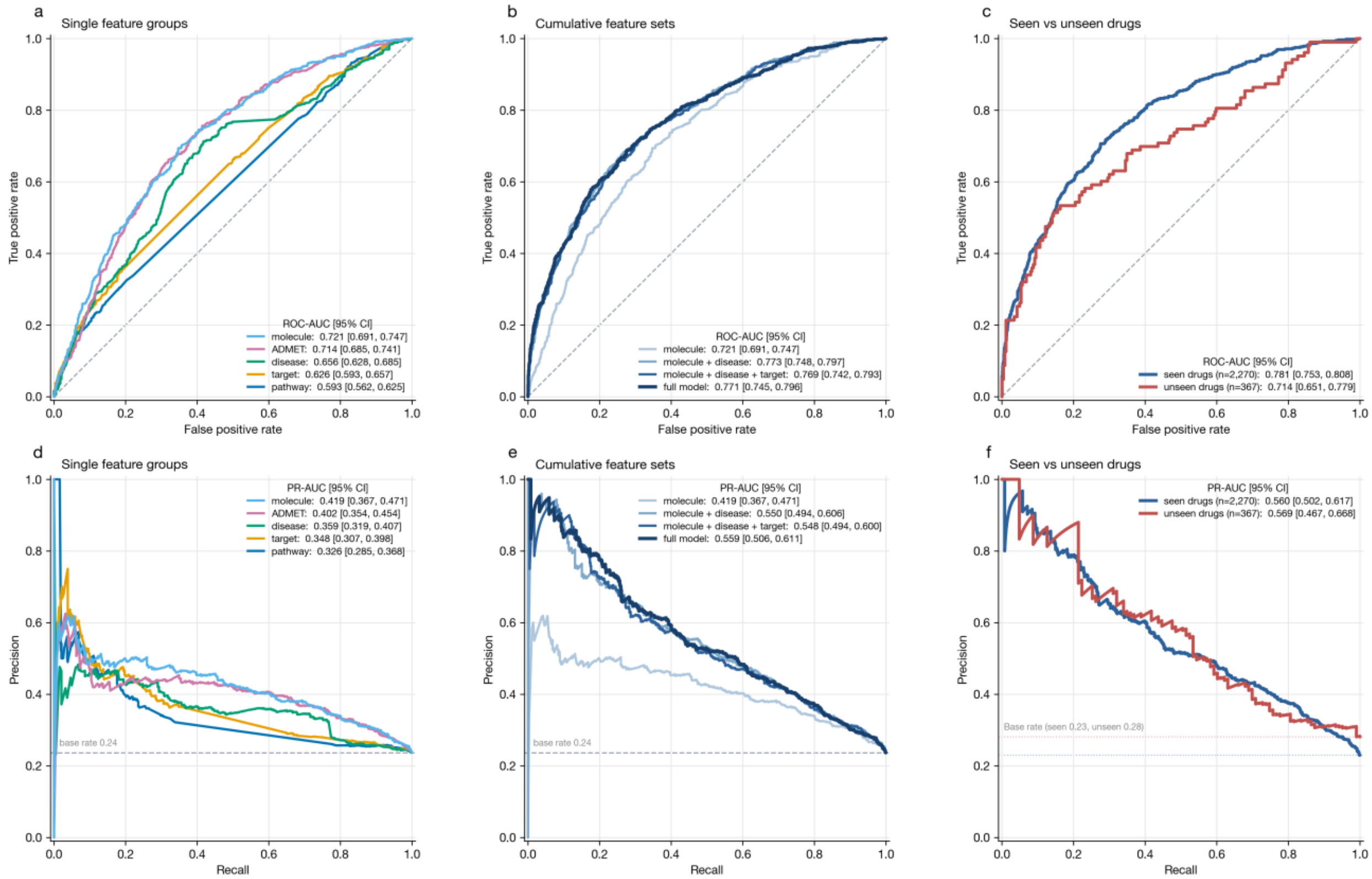

**Figure 4.** Feature-group signal and generalization across seen and unseen drugs. (**a, d**) ROC and precision-recall curves, respectively, for models trained on individual feature groups. (**b, e**) Corresponding curves for cumulative feature sets. (**c, f**) Full-model performance stratified by whether the drug appeared in training. Molecular and ADMET features provide the strongest individual signal. Adding disease identity to molecular features improves discrimination, further feature groups do not, and performance is lower for unseen than seen drugs. AUC values and 95% bootstrap confidence intervals are reported in the legends.

## Supplementary Methods

### *S1. Overview*

This document provides the full procedural detail for the dataset-construction pipeline, the labeling procedure, the feature representation, and the modeling layer summarized in the main-text Methods. Section numbering follows the order of the pipeline: data ingestion (S2 to S6), labeling (S7), cohort definition (S8 to S10), features (S11), models (S12 to S14), evaluation (S15), and the web tool (S16).

Throughout, a candidate is a unique drug-indication pair. A program is the set of clinical trials linked to a single candidate. A candidate is resolved when it is either approved for the matched indication by the study cutoff or classified as unapproved under the inactivity rule defined in S6. Candidates that do not meet either condition are unresolved and were excluded from the supervised dataset.

### *S2. Data sources and snapshot versions*

All data sources are public. Several are updated continuously. Where a resource could be pinned we recorded a fixed version or snapshot. AACT and the NLM ICD-10-CM Clinical Tables API were queried live and are identified by retrieval date instead. DrugBank, MeSH, and ICD-10-CM carry no usable version string, so they too are identified by retrieval date. These dates document reproducibility of the extracted feature set but do not imply that every annotation was available when each historical program began.

**Supplementary Table S1.** Data-source versions and retrieval dates.

| Resource | Version or snapshot date | Retrieved | Used for |
|---|---|---|---|
| ClinicalTrials.gov via AACT[1,2] | live database query (ctgov schema; no dated dump) | 2026-05-28 | Trial ingestion, program reconstruction |
| openFDA Drug Labeling API[3] | bulk dumps stamped 2026-04-15 to 2026-04-17; local SQLite mirror built 2026-05-08 | 2026-04-15 / 2026-05-08 | Indication-level approval labels |
| DrugBank[4] | Academic access; export file dated 2026-04-24 | 2026-04-24 | Drug name normalization, SMILES |
| ChEMBL[5] | CHEMBL36 | 2026-04-24 | Structure lookup, target annotations |
| Open Targets Platform[6,7] | 25.03 | 2026-05-28 | Target annotations, target-disease evidence |
| Reactome[8] | v96 | 2026-05-21 | Pathway annotations |
| MeSH[9] | no standalone version; from AACT browse_conditions and mesh_terms | 2026-05-28 | Condition normalization |
| ICD-10-CM (NLM Clinical Tables API)[10,11] | API v3, live endpoint, no edition stamp | 2026-05-28 | Disease coding |
| ADMET-AI[12] | 2.0.1 | n/a | ADMET property prediction |
| RDKit[13] | 2026.3.3 | n/a | Fingerprint generation |

### *S3. Clinical-trial ingestion*

We queried the AACT PostgreSQL mirror of ClinicalTrials.gov[1,2] and retained studies meeting all of the following criteria:

- Study type is interventional.
- At least one intervention of type “Drug”.
- At least one recorded condition.

Each retained study was expanded to one row per study × drug intervention × condition tuple, so a single trial testing two drugs across three conditions contributes six rows. We drew study identifiers, brief title, phase, overall status, reason for stopping, start and completion dates, last update submitted date, eligibility criteria text, intervention names, condition names, lead sponsor name, and MeSH intervention and condition terms from the AACT tables “studies”, “interventions”, “conditions”, “eligibilities”, “sponsors”, “browse_interventions”, “browse_conditions”, and “mesh_terms”. No date restriction is applied at ingestion. The year window described in S8 is applied later, to each candidate's earliest trial start date, after expansion and clustering.

We then removed rows matching either of the following:

Interventions whose name contained the string "placebo", matched case-insensitively against the intervention and indication fields. There is no regular expression and no match list, and vehicle controls are not separately identified.

Studies whose condition text matched a word-boundary regular expression for healthy-volunteer populations, covering "healthy" alone and "healthy" followed by volunteers, subjects, controls, participants, adults, individuals, males, females, men, women, or human.

Non-drug interventions never enter the expansion. Devices, procedures, dietary supplements, and behavioral interventions are excluded by the intervention_type = 'DRUG' join condition in the AACT query.

***S4. Drug normalization***

Intervention names in ClinicalTrials.gov are free text and include brand names, code names, dose strings, and formulation descriptors. We normalized each name in three stages.

Stage 1: string cleaning. We applied NFKD normalization and stripped diacritics, mapped µ to u, removed parenthetical, bracketed, and braced content, converted the separators / , _ ; : and hyphens to spaces, stripped quotes and remaining punctuation, normalized isotope labels so that "177 lu" becomes "177lu", and mapped the Roman numerals i to v onto 1 to 5. We then dropped a fixed list of dose, route, formulation, and trial-wording tokens, including injectable, injection, iv, oral, intravenous, subcutaneous, solution, powder, tablet, capsule, suspension, infusion, bolus, extended, release, dose, regimen, arm, placebo, sham, control, comparator, and standard.

Stage 2: dictionary matching. We matched cleaned names against DrugBank primary names, then the DrugBank synonym table, then the MeSH intervention leaf term, which was itself re-looked-up against DrugBank names and synonyms. Matching was exact after cleaning. No fuzzy or edit-distance matching is used anywhere in the pipeline.[4,9]

Stage 3: fallback. Names with no dictionary match were retained under a name-keyed identifier built from the cleaned string. These candidates carry no DrugBank identifier, and in practice they were dropped at the SMILES filter (S8) because no structure could be resolved for them.

Salt handling at the name stage is lexical rather than chemical. Eleven salt tokens (hcl, hydrochloride, sodium, potassium, acetate, sulfate, mesylate, tartrate, maleate, fumarate, phosphate) are stripped from the name. Chemical collapse to the parent structure happens later and structurally: SMILES strings were parsed with RDKit[13], standardized, reduced to the parent molecule by salt stripping and largest-fragment selection, and written back as canonical

SMILES. Parse and standardization failures are recorded as an explicit status rather than silently dropped.

### *S5. Condition normalization*

Condition strings were normalized to ICD-10-CM codes in two stages. The pipeline contains no curated MeSH to ICD-10 crosswalk and no UMLS concept mapping. Both stages issue free-text queries against the NLM ICD-10-CM Clinical Tables API.[9–11]

Stage 1: MeSH query. The most common MeSH leaf term from the AACT browse_conditions annotations was used as the free-text query string.

Stage 2: fallback query. When the MeSH query returned nothing, the cleaned raw condition string was used as the query instead. Queries and their results, including negative results, were cached locally.

Every code the API returned for a search term was accepted and stored, so a candidate may carry several codes. No similarity threshold, exact-match test, or top-ranked selection was applied. Codes were kept at the full billable granularity returned by the API and were never truncated to three-character categories.

Of 178,035 candidates, 88,583 received at least one ICD-10 code: 68,367 through the MeSH query and 20,216 through the raw-condition fallback. Conditions that could not be assigned a code were dropped at the ICD-10 filter (S8). This filter removes a substantial number of records and is the main reason the cohort under-represents rare and loosely specified conditions.

### *S6. Program reconstruction and trajectory matching*

Approval is a property of a development program, not of a single trial, so trials sharing a drug–indication pair must be linked before an outcome can be assigned. We followed the trajectory-matching logic described by Hay et al.[14]

For each candidate (drug–indication pair), we collected all trials sharing that pair and ordered them by phase and start date. A candidate is keyed by namespace and identifier joined to the normalized indication, for example db:DB11881__lymphoma, with the namespace drawn from DrugBank, MeSH, or the cleaned free-text name. A program was defined as the full ordered set of trials sharing a candidate key. We then assigned each program a terminal state using the following rules, evaluated in order: Outcomes were assessed as of 2026-05-07.

Approved. The adjudicator described in S7 matched the candidate indication to at least one of the drug's FDA-approved label indications, and the recorded approval date falls on or before the pipeline's as-of date. Approval is not tested against the start date of the earliest trial in the program.

Unapproved. No approved label indication matched, and the candidate's most recent ClinicalTrials.gov update was submitted more than two years before the as-of date. This is an inactivity rule on the last update date. It is not a rule on AACT terminal status, and it is not a rule on trial start dates. The operational state is recorded together with the phase reached. This inactivity rule identifies resolved unapproved programs and should not be interpreted as a uniform clinical efficacy failure.

Unresolved. Neither condition holds, for example because the candidate's most recent trial update falls inside the two-year window. The pipeline emits these candidates as Ongoing or Unknown. Both classes were excluded from the supervised dataset.

Phase gaps were treated as follows. A program is not required to contain trials from every phase, and a program consisting of a single Phase 1 trial is eligible for either terminal state. Trials with missing or unrecognized phase are mapped to Unknown and are ignored for cohort and advancement purposes, as are trials marked not applicable. Combined phases map upward, so Phase 1/2 becomes Phase 2 and Phase 2/3 becomes Phase 3.

Right-censoring. The pipeline applies no censoring rule based on trial start dates. Recency enters only through the two-year inactivity window described above, which determines whether an unapproved candidate is classified as unapproved or is left unresolved and excluded. We state this explicitly because it directly affects the observed approval rate. See S10.

### *S7. Approval labeling*

Indication-level approval status is not available as a single structured public field. FDA labels state approved indications in unstructured text, and trial conditions are recorded as separate free-text strings. We therefore adjudicated the match between the two with a locally hosted language model.

#### *S7.1 Label extraction*

For each drug in the cohort, we retrieved all label records from a local SQLite mirror of the openFDA drug label, NDC, and Drugs@FDA bulk dumps, and extracted the "indications_and_usage" field. Where a drug had several label records, whether from different manufacturers or from successive revisions, we took the union of the stated indications across all matching records, deduplicating by exact string equality. Label effective dates are stored but are not used to select a revision. Label text was split into indication strings on numbered section markers, with a sentence-boundary fallback, and normalized before deduplication. Where more than twenty indications remained, we ranked them by token-set similarity against the trial and MeSH indication strings and kept the top twenty.[3]

### *S7.2 Language-model adjudication*

We used Qwen 2.5-7B-Instruct[15] to judge whether a trial indication is covered by the drug's approved-label indication text. The model was given the trial indication string, the MeSH indication string, the matched drug synonym, and the numbered list of approved-label indications. It returned a binary judgment, a confidence score, the verbatim text of the matched label indication, and a brief justification. No ICD-10 code and no ICD-10 description is passed to the model. No remote or commercial model is used at this stage.

Decoding used temperature 0 and a maximum of 256 output tokens, with JSON object response formatting. No top-p value, no seed, and no token-count override was set. The exact prompt is reproduced in Supplementary Note 1.

### *S7.3 Label validation*

We assessed the quality of the model-extracted approval labels by manually adjudicating a stratified sample of 100 candidates: 50 labeled "approved" and 50 labeled "not approved" by the pipeline. For each candidate, the trial indication was manually compared with the exact FDA indication text supplied to the model, using the matching criteria defined in S7.2.

Manual review confirmed 48 of the 50 pipeline “approved” decisions, yielding a positive predictive value (PPV) of 96.0%. Among the 50 pipeline “not approved” decisions, 46 were confirmed, yielding a negative predictive value (NPV) of 92.0%. Because the sample was stratified by pipeline-assigned label rather than natural class balance, PPV and NPV are reported separately rather than as a pooled overall accuracy estimate.

| Drug | Trial indication | Pipeline decision | Manual agreement | Basis for manual adjudication |
|---|---|---|---|---|
| Cyclophosphamide | Recurrent transformed non-Hodgkin lymphoma | Approved | Agree | FDA prescribing information includes malignant lymphomas among the indicated malignancies. The trial indication is explicitly a non-Hodgkin lymphoma; transformed follicular lymphoma is also described as an aggressive non-Hodgkin lymphoma.[16,17] |
| Oxaliplatin | Unresectable pancreatic cancer | Not approved | Agree | FDA prescribing information limits oxaliplatin's indicated uses to adjuvant treatment of stage III colon cancer and treatment of advanced colorectal cancer; pancreatic cancer is not included.[18] |
| Pazopanib | Uterine leiomyosarcoma | Not approved | Disagree | FDA prescribing information includes advanced soft tissue sarcoma following prior chemotherapy. Leiomyosarcoma is a recognized subtype of soft tissue sarcoma, including uterine leiomyosarcoma, and is therefore covered by the broader labeled category.[19,20] |

***S8. Cohort filtering and attrition***

The table below reports record counts at each stage of the pipeline. Counts are of unique drug–indication candidates unless otherwise stated. The rows follow the order in which the filters are applied, which places the SMILES filter before the removal of unresolved programs. Because the filters are set intersections, the final count does not depend on their order.

**Supplementary Table S2.** Cohort attrition by pipeline stage.

| Stage | Records remaining | Removed at this stage |
|---|---|---|
| AACT interventional studies retrieved | 196,291 studies | n/a |
| Expanded to study x drug x condition tuples | 679,338 | n/a |
| After removing placebo and healthy-volunteer rows | 596,300 | 83,038 |
| Collapsed to unique drug-indication candidates | 178,035 | n/a |
| After restriction to post-2009 programs | 144,731 | 33,304 |
| After SMILES availability filter | 55,350 | 89,381 |
| After removing unresolved (Ongoing or Unknown) programs | 31,977 | 23,373 |
| After ICD-10 availability filter | 14,134 | 17,843 |
| Of which approved / unapproved | 2,469 / 11,665 | n/a |

Two counts require qualification. The 196,291 studies are those surviving the ingestion join, that is, interventional studies with at least one drug intervention and at least one recorded condition; a count of all interventional studies before the drug restriction is not persisted. Placebo and healthy-volunteer rows are removed in a single pass and only their union is logged.

The SMILES and ICD-10 filters carry the most consequence for interpretation. The SMILES requirement removes biologics, cell and gene therapies, mixtures, and other modalities that cannot be represented as a single small-molecule structure. The ICD-10 requirement removes conditions expressed as symptom clusters, composite endpoints, or non-standard trial-specific descriptions. Both filters shift the cohort toward well-characterized small molecules developed for well-coded diseases.

### *S9. Temporal split*

Programs beginning on or before 31 December 2019 formed the training set; programs beginning after that date formed the held-out test set. The split is by program start date, defined as the start date of the earliest trial in the program, so that no candidate contributes trials to both sides of the boundary.

The training set contains 11,497 candidates, of which 1,845 (16.05%) are approved. The test set contains 2,637 candidates, of which 624 (23.66%) are approved. No candidate was dropped for a missing start year.

We chose a temporal rather than a random split because a random split would allow the model to see later trials of a program whose earlier trials appear in training, and because the intended use of the model is prospective. A drug may appear in both splits when it is developed for different indications at different times. We quantify the effect of this overlap in the seen-versus-unseen analysis (S15.3).

A single stratified 10% slice of the training set was held out. Encoders, scalers, dimensionality-reduction transforms, calibration models, and classifier parameters were fit only on the remaining 90%; no component was fit on the held-out post-2019 test set.

### *S10. Base rate and survivorship disclosure*

The observed approval rate in our cohort is 17.5% (2,469 of 14,134), which is higher than the roughly 10% to 14% probability of approval from Phase 1 reported in the LOA literature[14,21]. Three cohort-construction choices account for most of this gap, and we state them here so that model scores are not read as calibrated field-wide approval probabilities.

Modality restriction. The SMILES requirement retains small molecules and removes modalities with different, and often lower, success profiles. It also biases molecule inclusion toward molecules with better documented structures which may in turn filter out earlier experimental molecules more frequently than approved molecules.

Program resolution requirement. Candidates whose programs remain active are excluded. An unapproved candidate is classified as unapproved only after two years of inactivity on ClinicalTrials.gov (S6), so recently abandoned programs are held out as unresolved rather than counted as resolved negatives. Because non-approval or discontinuation can be recorded sooner than approval, we report the observed rate under this rule rather than adjusting it.

Registration threshold. Our unit of analysis begins at the first registered trial, so candidates abandoned before any trial was registered do not appear. Preclinical attrition is therefore invisible to the cohort by construction.

Outcome-maturity sensitivity. Because unresolved programs are excluded, later test years are increasingly enriched for rapidly resolving programs. We therefore repeated evaluation on the 1,486 test programs beginning in 2020–2021 (14.1% approved). ROC-AUC was 0.79 [0.75, 0.82] for the full model, 0.80 [0.76, 0.83] for molecule + disease, 0.73 [0.69, 0.77] for molecule only, 0.62 [0.58, 0.67] for adapted HINT, and 0.46 [0.42, 0.51] for ChemAP. Full-model PR-AUC decreased from 0.56 on the full test set to 0.44 in this subset as approval prevalence decreased from 23.7% to 14.1%. Intermediate restrictions through 2022, 2023, and 2024 showed the same gradual pattern. Across test start years, the observed approval rate rose from 11.8% for 2020 starts to 100% for 2026 starts, while ongoing exclusions increased from approximately 46% to 89%.

The comparison below places our cohort next to published field rates. These are not directly comparable, because the denominators differ, and we present them only to make the magnitude of the offset explicit. The rates attributed to Hay et al. and Wong et al. are quoted from those papers and were not recomputed here.[14,21]

| Source | Unit | Reported approval rate | Denominator |
|---|---|---|---|
| This work | Resolved drug-indication candidate | 17.5% | Registered, resolved, SMILES- and ICD-10-codable programs |
| Hay et al.[14] | Drug-indication program | 10.4% | Industry-curated development histories |
| Wong et al.[21] | Drug-indication path | 13.8% | ClinicalTrials.gov program histories |

Calibration improves held-out probability agreement but does not fully eliminate temporal base-rate shift. Users of the web tool should interpret scores cautiously with the expectation that probabilities may be overestimated for some inputs and underestimated for others. We recommend using the tool as a decision support system rather than a decision gate.

### *S11. Feature representation*

Features fall into five groups. Every group is computed from the drug, the disease, or both, and none uses trial protocol information, sponsor identity, phase, or trial results. This is what we mean by trial-agnostic. We treat molecule + disease as the cleanest pretrial feature set; ADMET, target, and pathway features are richer contemporary annotations that could plausibly be available before trials.

### *S11.1 Molecule*

We generated ECFP4 Morgan fingerprints with radius 2 and 2,048 bits, without chirality or feature invariants, together with 167-bit MACCS keys, using RDKit[13]. In our generated feature matrix, the MACCS block contained 167 columns. Fingerprints were computed on the standardized parent structure described in S4. The molecule group has a raw dimension of 2,215. No scalar physicochemical descriptors were included. Molecular weight, logP, topological polar surface area, and rotatable-bond counts do not enter the molecule group, and physicochemical quantities reach the model only through the ADMET percentiles described in S11.3.

### *S11.2 Disease*

Each candidate's ICD-10 codes were expanded to the code itself and its full ancestor chain, giving four tokens per code: the full dotted code, the three-character category, the block, and the chapter. The code E74.02, for example, yields ICD:E74.02, ICD3:E74, BLOCK:E70-E88, and CHAPTER:IV. These tokens were multi-hot encoded with no vocabulary cap, giving a raw dimension of 7,175: 7,173 tokens, a residual-count feature, and a missingness indicator.[10,11]

### *S11.3 ADMET*

We ran ADMET-AI 2.0.1[12] on each standardized structure; the installed version returned 52 predicted properties spanning absorption, distribution, metabolism, excretion, and toxicity endpoints. We used ADMET-AI's percentile outputs rather than its raw predictions. These percentiles are computed by ADMET-AI against its own fixed reference set rather than our training distribution, so the transform is external to our supervised labels and was not fit on this dataset. Because ADMET-AI is a contemporary pretrained model, we do not treat these features as historically leakage-protected. What is fit on the inner training slice and applied unchanged to the validation and test sets is the per-column median imputation of missing values and the PCA described in S11.6. No column was dropped and no missingness indicator was required, because every ADMET column had fewer than 5% null values on the training data.

### *S11.4 Target*

Target annotations were drawn from ChEMBL and Open Targets[5–7.] The locked model uses target availability scalars (a target count and a data-availability flag), a mechanism-of-action multi-hot encoding capped at the 200 most frequent mechanisms, an action-type multi-hot encoding over the 20 observed types, five per-(target, indication) genetic-support scalars (the maximum and mean genetic score for the indication, the number of targets with genetic support, and flags for any genetic support and for an indication–target match), three LOEUF constraint scalars (the minimum and mean LOEUF across targets and a data-availability flag), and tractability multi-hot encodings over four modalities and nineteen labels. The raw target dimension is 271.

Drugs with no annotated target receive all-zero vectors together with explicit missingness flags, including the target-data flag, the mechanism and action-type missingness indicators, the two tractability missingness indicators, and the flags for genetic support, indication–target match, and LOEUF availability. Missingness is itself informative, so these flags were retained rather than suppressed. 34.2% of the cohort has no target annotation, and the rate is higher in the test set (42.4%) than in the training set (32.4%), because more recent programs are less completely annotated.

### *S11.5 Pathway*

Pathways were assigned by mapping each drug's targets to Reactome pathways through Open Targets[6–8]. The pathway group has a raw dimension of 1,464 and comprises six Reactome context scalars with a missingness indicator, a multi-hot encoding of top-level pathway identifiers over a vocabulary of 29, a multi-hot encoding of ancestor pathway identifiers capped at 500, a weighted map of ancestor pathway scores over 414 pathways, a multi-hot encoding of best-supported target pathway identifiers capped at 500, and a count of pathways reached from supported targets. Both 500-item caps are reached. Reactome enrichment returned at least one pathway for 54,616 of 178,035 candidates (31%).

Two issues required specific handling.

Ancestor-chain redundancy. For Reactome pathways[8], we encoded each annotated pathway together with its available ancestor chain. We walked the full chain of strict ancestors upward to the roots, rather than using leaf pathways alone or truncating at a fixed level. In our encoding, distinct lower-level pathways can share signal through common higher-level ancestors. This is the same rationale as the ICD-10 ancestor expansion in S11.2. Roots are captured separately by the top-level pathway encoding, and the local leaf count is retained as its own feature. Candidates with no pathway annotation, 39.2% of the cohort (37.6% of the training set and 45.8% of the test set), receive all-zero vectors and a missingness indicator.

Feature timing. Target and pathway annotations were taken from current public resources which represent a more comprehensive state of information than what may have been known at the time of the clinical trial. However, this is information that is not necessarily dependent on clinical trial outcomes and can be known before clinical trials begin. We therefore treat this as a plausible prediction scenario but nonetheless found that adding target and pathway information generally did not meaningfully shift model performance (Table S3).

### *S11.6 Dimensionality reduction and missing values*

The cohort is small relative to the raw feature dimensionality, so we compressed each feature group to 50 components, or to $d - 1$ components where the group's raw dimension $d$ is smaller. Dense groups were reduced with PCA. The disease group, a large sparse multi-hot encoding, was reduced with truncated SVD, which does not require centering and preserves sparsity. The target and pathway groups are composite blocks of scalars, flags, and multi-hot encodings, and are reduced with PCA rather than SVD.

| Feature group | Raw dimension | Reduction | Final dimension |
|---|---|---|---|
| Molecule (ECFP4 + MACCS) | 2,215 | PCA | 50 |
| Disease (ICD-10 multi-hot) | 7,175 | Truncated SVD | 50 |
| ADMET | 52 | PCA | 50 |
| Target | 271 | PCA | 50 |
| Pathway | 1,464 | PCA | 50 |
| Total model input | 11,177 | - | 250 |

All reduction transforms were fit on the inner training slice only and applied unchanged to the validation and test sets. Missing values were handled per group. Unparseable SMILES yield an all-zero bit vector. Candidates with no ICD-10 code receive an all-zero multi-hot vector and a missingness flag. ADMET columns are median-imputed with medians computed on the inner training slice. Target and pathway groups receive all-zero vectors together with the explicit availability flags described above, and their scalar components are median-imputed alongside paired indicators.

We verified that no protocol-derived field enters the feature space. The encoder map reads only the SMILES string, the ICD-10 codes, and the ADMET, ChEMBL, Open Targets, and Reactome columns. Trial phase is present in the schema but is used only to stratify evaluation, never as an input. Sponsor identity, eligibility criteria, and trial results never enter the model.

### *S12. Model and training*

The primary classifier is a gradient-boosted decision-tree model implemented with XGBoost 3.2.0[22]. Hyperparameters were fixed across all experiments so that differences between rows of main-text Table 2 reflect feature sets and task framing rather than per-model tuning. XGBoost used a binary logistic objective, up to 1,000 trees with maximum depth 6, a learning rate of 0.05, row and column subsampling of 0.85, class weighting based on the training-set negative-to-positive ratio (5.23), and early stopping after 50 rounds using validation ROC-AUC. Training used the histogram tree method and random seed 0. Encoder vocabularies (ICD-10 tokens, mechanisms, pathways) are ordered by training-set frequency with a deterministic tie-break, so training is reproducible across runs.

Class imbalance was addressed through scale_pos_weight rather than resampling; no oversampling or undersampling was applied. The value follows from the trainingsplit, which contains 9,652 unapproved and 1,845 approved candidates. A single stratified 10% slice of the training set was held out. Encoders, dimensionality-reduction transforms, and the classifier were fit on the remaining 90%, while the held-out slice was used solely for gradient-boosting early stopping. Calibration of the served model is described in S13. Hyperparameters were fixed a priori, so no model selection was performed on the held-out slice. No component was refit on the complete training set, and the post-2019 test set remained untouched by all fitting.

### *S13. Calibration*

Calibration was applied to the molecule-and-disease model served by the web tool, while model performance reported in the paper was generally computed from raw, uncalibrated scores. Discrimination metrics are invariant to monotone rescaling, so calibration does not affect the reported ROC-AUC or PR-AUC.

For the molecule-and-disease model, raw scores were mapped to probabilities with sigmoid calibration (Platt scaling)23, implemented as a logistic regression with C = 1e6 fit on the one-dimensional model score. The calibrator was fit on out-of-fold predictions from five-fold cross-validation over the training set: each fold was scored by a model trained on the other four (each with its own 10% early-stopping slice), so no candidate contributed to fitting the model that scored it. It was then applied to test-set scores from the model fit on the training split. Neither the classifier nor any preprocessing component was refit on the complete training set.

Recent test programs are right-censored - a program starting in 2025 appears only if it has already resolved, and such programs are predominantly approvals (observed approval rate 0.55, 0.96 and 1.00 for 2024–2026 starts, against a flat 0.14–0.19 across all development years) - so probability metrics were assessed on the 1,486 test programs starting 2020–2021. Calibration lowered the Brier score from 0.12 to 0.10 and expected calibration error, over ten equal-frequency bins, from 0.14 to 0.02. Over the full test set the corresponding values are 0.15 and 0.15, reflecting the censoring-inflated base rate rather than miscalibration. The reliability diagram is Supplementary Figure S1.

### *S14. Baseline adaptation*

#### *S14.1 HINT*

HINT[24] predicts trial outcome from molecule, disease, and eligibility criteria as well as, optionally, ADMET/PK descriptors. Two adaptations were needed to compare our model results to a HINT-style model.

First, eligibility criteria are protocol-derived and unavailable at our target decision point, so we disabled that input. The protocol-embedding module was retained in the architecture; what was removed was its input. Criteria strings were blanked, and HINT's sentence-embedding table was repointed at an empty dictionary, so the module pools nothing and emits a constant zero vector. Second, we retrained the adapted architecture on our indication-level labels using our temporal split, with the validation fold carved from the training split with stratification by label. Training used 5 epochs, a learning rate of 1e-3, Adam without weight decay, unweighted binary cross-entropy, a batch size of 32, a 10% validation fraction, seed 42, and CPU execution. Other architectural settings followed the published HINT architecture.[24]

To check that the adapted model's weak performance reflects task mismatch rather than a broken reimplementation, we also ran the HINT implementation[24] on the TOP Phase 1, Phase 2, and Phase 3 benchmark files used in this evaluation alongside a simple XGBoost comparator built from the same inputs. The benchmark files used here contained 627, 1,654, and 1,146 rows and 55.3%, 55.5%, and 75.0% positive. Those results appear in Supplementary Table S4.

### *S14.2 ChemAP*

ChemAP predicts drug-level FDA approval from chemical structure[25]. We applied the released model out of the box, with no retraining, using the upstream repository at commit 3f057858ecf756035b507de403f518c3c1d66996 and the released DrugApp-trained student checkpoints for seed 7 in the knowledge-distilled variant. The prediction is the mean of the ECFP and SMILES student probabilities. Our vendored copy drops the torch_geometric dependency and is otherwise unmodified. ChemAP reads only the first SMILES string per row and ignores disease, criteria, and phase. Each drug in our test set therefore received one score, which was compared against every drug–indication label for that drug, so a drug with four test-set indications contributes its single score four times.

We did not retrain ChemAP, because the point of the comparison is to ask whether an existing drug-level approval score transfers to the indication-level task, not whether the ChemAP architecture could learn our task.

### *S14.3 Perfect drug-level oracle*

To separate task mismatch from ChemAP-specific model error, we constructed an oracle with perfect knowledge of drug-level approval status but no indication resolution. For each drug, the oracle takes the maximum label over all of that drug's candidates in the whole dataset, train and test together, and assigns that value to every one of the drug's test candidates. Drugs absent from the lookup receive zero. The oracle therefore reads our own labels, including test labels. That is deliberate, and it is what makes the comparison a ceiling rather than a baseline: the drug-level knowledge comes from the labels themselves. Because the oracle emits only 0 and 1, ties are pervasive. The area under the curve was computed analytically with the rank-based estimator, which scores tied pairs at exactly 0.5, rather than by breaking ties at random.

This oracle achieves ROC-AUC 0.71 and PR-AUC 0.35, well below our task-specific model. This result places a ceiling on drug-level transfer and establishes that the gap in main-text Figure 3 is a property of the prediction unit rather than of ChemAP.

### *S15. Evaluation*

#### *S15.1 Metrics*

ROC-AUC and PR-AUC are the primary discrimination metrics. PR-AUC is reported because the positive class is a minority (23.7% of the test set) and PR-AUC is more sensitive than ROC-AUC to performance on the positive class.

For reference, we also report a test-set-optimized maximum F1: the precision-recall curve is swept over the held-out test set and the best operating point is taken. For the five-group model this exploratory maximum is F1 0.54 at threshold 0.35 and should not be interpreted as performance at a pre-specified operating threshold. Fixed-threshold summaries are reported separately at threshold 0.5: 250 true positives, 180 false positives, 1,833 true negatives, and 374 false negatives, giving accuracy 0.79, precision 0.58, recall 0.40, F1 0.47, and balanced accuracy 0.66. Threshold-dependent summaries are collected in Supplementary Table S3.

#### *S15.2 Confidence intervals*

ROC-AUC and PR-AUC confidence intervals for all model comparisons were computed with 1,000 drug-cluster bootstrap replicates, resampling drugs with replacement and retaining all associated candidate rows. Replicates containing a single class were skipped, and 95% percentile intervals are reported. For the full model, ROC-AUC was 0.77 [0.75, 0.80] and PR-AUC was 0.56 [0.51, 0.61]. F1 intervals were computed by candidate-level bootstrap resampling.

Paired model differences used the same drug-cluster bootstrap design. For the full versus molecule + disease comparison, ΔROC-AUC was −0.002 [−0.019, +0.014] and ΔPR-AUC +0.009 [−0.015, +0.033]; neither interval excludes zero.

### *S15.3 Seen versus unseen drugs*

We split the test set by whether the drug appeared anywhere in the model's training or internal validation rows, regardless of indication. Drugs are identified by their DrugBank identifier. The unseen subset contains 367 candidates, of which 103 are approved; the seen subset contains 2,270 candidates, of which 521 are approved. Because a drug may be seen through a different indication, the seen stratum does not imply that the model has seen the label being predicted. The five-group model reaches ROC-AUC 0.78 [0.75, 0.81] on seen drugs and 0.71 [0.65, 0.78] on unseen drugs. The corresponding score distributions are shown in Supplementary Figure S2.

### *S15.4 Within-drug AUC*

Disease-selection performance was evaluated across 212 drugs with at least one approved and one unapproved indication, yielding 1,721 approved–unapproved indication pairs. The full model achieved within-drug AUC 0.71 [0.66, 0.76], compared with 0.67 [0.63, 0.71] for disease only, 0.71 [0.65, 0.76] for molecule + disease, and 0.50 for molecule only. In the top-1 analysis, the full model selected an approved indication for 61.3% [54.7%, 67.9%] of drugs, compared with 50.9% [44.3%, 58.0%] for disease only, 59.9% [53.8%, 66.5%] for molecule + disease, and 40.7% [38.1%, 43.1%] under random selection. Confidence intervals and paired differences were estimated by drug-level bootstrap resampling.

***S16. Web tool***

The web tool serves the calibrated molecule-and-disease model, not the five-group model whose performance is reported in the main text. The served model reaches ROC-AUC 0.77 on the held-out test set, matching the 0.77 of the five-group model. The deployed artifact bundles the fitted encoders, the PCA and SVD transforms, the classifier, and the Platt calibrator, so the preprocessing applied at inference is exactly the preprocessing used when that model was trained.

A request supplies a SMILES string and an ICD-10 code, or a drug name and a disease name. Drug names are resolved to SMILES using ChEMBL, and disease names to ICD-10-CM codes through the NLM Clinical Tables API[5,10,11]. The service standardizes the structure exactly as in S4, computes the molecule and disease feature groups, applies the fitted reduction transforms, scores the candidate, and applies the Platt calibrator. Prediction failures return HTTP 400 and resolver failures return HTTP 502. Structure parsing and ICD-10 vocabulary coverage are surfaced as diagnostic flags rather than as errors: an unparseable SMILES string yields an all-zero fingerprint and a scored prediction carrying a validity flag, not a refusal.

When a query exactly matches a molecule–disease pair present in the model-development data, the service returns the observed label (1.0 or 0.0) as a known-outcome lookup rather than asking the model to predict an already observed result. Other queries receive the calibrated model prediction. This lookup behavior is part of the translational interface only and is not used in any reported model-performance evaluation.

The tool is implemented with FastAPI and uvicorn behind a single-page HTML front end, containerized, and deployed to Hugging Face Spaces at https://huggingface.co/spaces/samirkt/drug-disease-approval-forecasting. User inputs are not logged or persisted; the service writes only a startup line recording that the model has loaded.

## Supplementary Notes

**Supplementary Note 1.** Full language-model prompt for indication adjudication, including the system prompt, the user template, and the output schema. The prompt is zero-shot and contains no worked examples.

# System prompt

You are a regulatory affairs analyst. Decide whether a clinical-trial
indication for a specific drug is covered by any of that drug's
FDA-approved label indications.

INPUTS:
- TRIAL INDICATION: free-text disease name from a clinical trial.
- APPROVED LABEL INDICATIONS: numbered list taken verbatim from the drug's current FDA label(s). MAY contain unrelated indications.

RULES:
- approved=true ONLY if at least one label indication substantively covers the trial indication. A broader approved indication that encompasses the trial indication counts. A narrower approved indication does NOT cover a broader trial indication.
- approved=false if none match, or you cannot tell with reasonable confidence. Default to false on uncertainty.
- matched_indication: copy the verbatim text of the matched label indication when approved=true; null otherwise. Do NOT invent text.
- confidence: 0.0-1.0.
- reasoning: 1-3 sentences.

CRITICAL: Do NOT reason about the drug's known indications from your
training data. Use only the label list provided.

# User template

TRIAL INDICATION: {trial_indication}
MESH INDICATION: {mesh_indication}

APPROVED LABEL INDICATIONS for {matched_synonym}:
{numbered_label_indications}

# Required output schema (JSON)

```
{
  "type": "object",
```

```
  "properties": {
    "approved": {
      "type": "boolean"
    },
    "confidence": {
      "type": "number",
      "minimum": 0.0,
      "maximum": 1.0
    },
    "matched_indication": {
      "type": [
        "string",
        "null"
      ]
    },
    "reasoning": {
      "type": "string"
    }
  },
  "required": [
    "approved",
    "confidence",
    "matched_indication",
    "reasoning"
  ]
}
```

## Supplementary Tables

### Supplementary Table S3. Threshold-dependent metrics at threshold 0.5 with 95% percentile bootstrap CIs (1000 replicates); max-F1 for reference.

| Model | n | Accuracy | Precision | Recall | F1 | Balanced acc. | max-F1 (thr) |
|---|---|---|---|---|---|---|---|
| ChemAP (structure-only, transferred) | 2637 | 0.280 [0.264, 0.297] | 0.224 [0.208, 0.242] | 0.832 [0.800, 0.861] | 0.353 [0.331, 0.376] | 0.470 [0.454, 0.486] | 0.383 (0.004) |
| HINT (adapted, molecule + disease) | 2637 | 0.512 [0.495, 0.532] | 0.282 [0.262, 0.305] | 0.688 [0.654, 0.725] | 0.400 [0.377, 0.426] | 0.573 [0.553, 0.594] | 0.405 (0.483) |
| molecule | 2637 | 0.686 [0.667, 0.702] | 0.395 [0.364, 0.425] | 0.617 [0.579, 0.654] | 0.482 [0.450, 0.510] | 0.662 [0.641, 0.683] | 0.493 (0.460) |
| disease | 2637 | 0.666 [0.647, 0.684] | 0.349 [0.318, 0.381] | 0.476 [0.437, 0.516] | 0.402 [0.371, 0.434] | 0.600 [0.578, 0.623] | 0.466 (0.434) |
| ADMET | 2637 | 0.675 [0.657, 0.693] | 0.390 [0.361, 0.418] | 0.659 [0.622, 0.696] | 0.490 [0.461, 0.518] | 0.669 [0.649, 0.689] | 0.491 (0.500) |
| target | 2637 | 0.706 [0.688, 0.722] | 0.364 [0.322, 0.403] | 0.327 [0.290, 0.364] | 0.345 [0.310, 0.380] | 0.575 [0.555, 0.595] | 0.410 (0.332) |
| pathway | 2637 | 0.714 [0.697, 0.731] | 0.359 [0.317, 0.402] | 0.264 [0.233, 0.300] | 0.305 [0.270, 0.340] | 0.559 [0.541, 0.577] | 0.404 (0.473) |
| molecule + disease | 2637 | 0.785 [0.768, 0.800] | 0.557 [0.513, 0.598] | 0.439 [0.399, 0.476] | 0.491 [0.453, 0.525] | 0.665 [0.644, 0.685] | 0.538 (0.359) |
| molecule + disease + target | 2637 | 0.787 [0.771, 0.802] | 0.573 [0.525, 0.617] | 0.391 [0.349, 0.431] | 0.465 [0.424, 0.502] | 0.650 [0.628, 0.671] | 0.533 (0.309) |
| full (molecule + disease + target + pathway + ADMET) | 2637 | 0.790 [0.774, 0.805] | 0.581 [0.537, 0.624] | 0.401 [0.361, 0.440] | 0.474 [0.436, 0.509] | 0.656 [0.635, 0.675] | 0.539 (0.347) |

**Supplementary Table S4. TOP Phase 1/2/3 benchmark comparison: HINT[24] vs XGBoost comparator (molecule + disease inputs).**

| Phase | Model | n | positive_frac | roc_auc | pr_auc | f1 |
|---|---|---|---|---|---|---|
| Phase 1 | HINT | 627 | 0.553 | 0.548 | 0.615 | 0.713 |
| Phase 1 | XGBoost | 627 | 0.553 | 0.565 | 0.596 | 0.713 |
| Phase 2 | HINT | 1654 | 0.555 | 0.616 | 0.665 | 0.714 |
| Phase 2 | XGBoost | 1654 | 0.555 | 0.613 | 0.665 | 0.716 |
| Phase 3 | HINT | 1146 | 0.750 | 0.688 | 0.861 | 0.857 |
| Phase 3 | XGBoost | 1146 | 0.750 | 0.702 | 0.862 | 0.857 |

**Supplementary Figures**

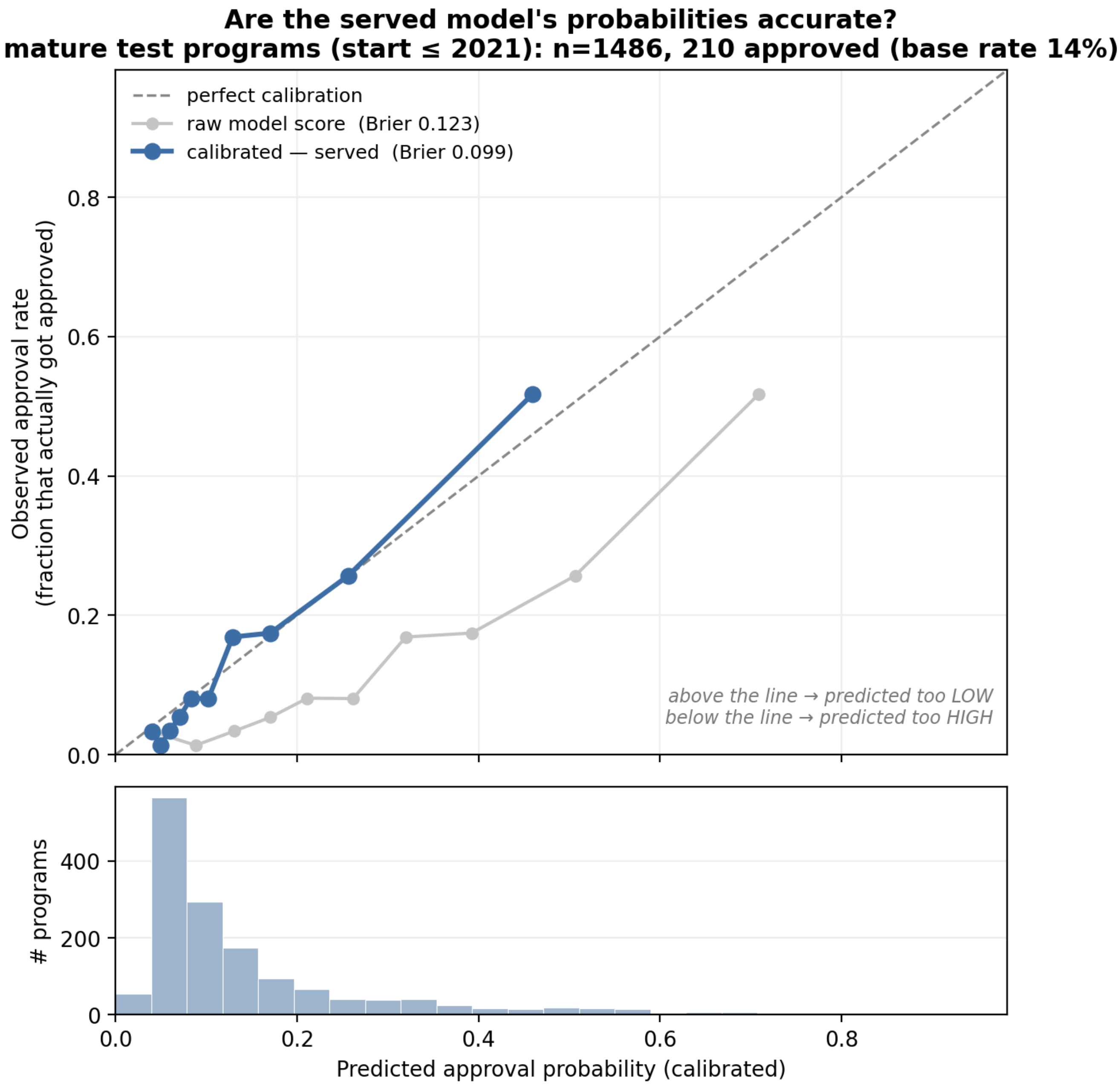

**Supplementary Figure S1.** Reliability diagram and Brier score for the served molecule-and-disease model on the 1,486 held-out test programs starting 2020–2021.

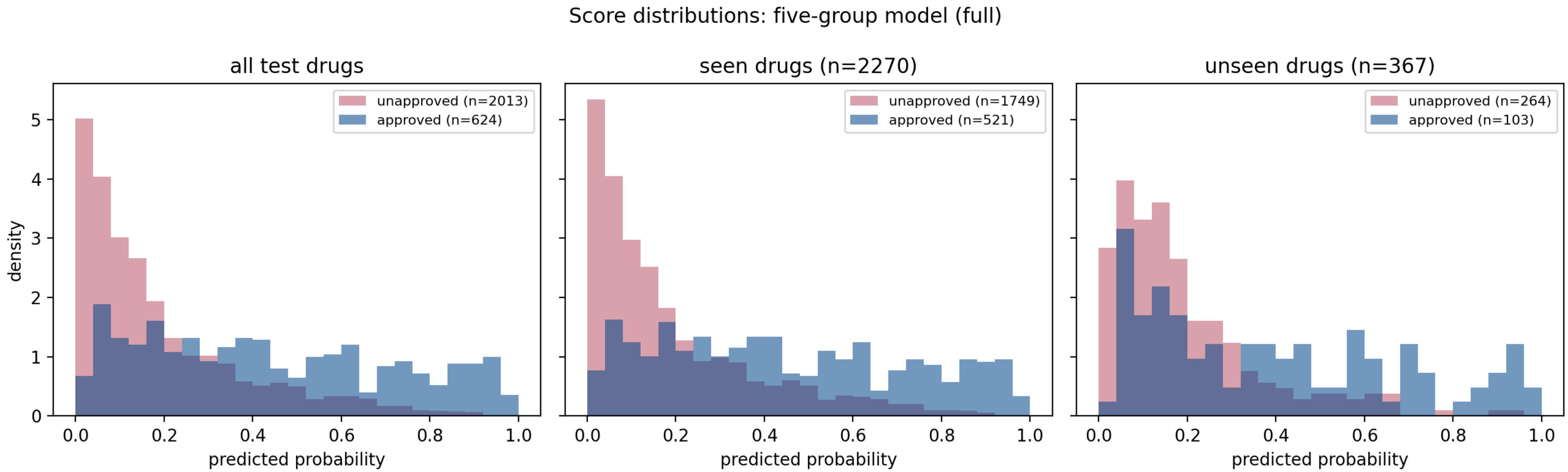

**Supplementary Figure S2.** Score distributions for approved and unapproved candidates, overall and stratified by seen versus unseen drugs.